\documentclass[aps,prl,preprint,tightenlines,superscriptaddress,showpacs,byrevtex]{revtex4}
\usepackage{graphicx} % Include figure files
\usepackage{dcolumn}  % Align table columns on decimal point

\graphicspath{{ps}}

\begin{document}

%\vspace*{-3\baselineskip}
%\resizebox{!}{3cm}{\includegraphics{belle.eps}}

\preprint{\vbox{ \hbox{   }
                 \hbox{BELLE-CONF-0441}
                 \hbox{ICHEP04 11-0687} 
                 %KM \hbox{hep-ex nnnn, if available}
}}

\title{ \quad\\[0.5cm] Measurement of Branching Fractions 
in  $B^0 \rightarrow J/\psi \pi^+ \pi^-$ Decay}

%%%% >>>>> insert the authorlist here. BEFORE the abstract !!!!! <<<<<
%\input{author-conf2004.tex}
%%% Paper:    
%%% Journal:  summer 2004 conference papers (PRL format)
%%% Contacts: 
%%% Last revised on July 14, 2004 16:40:00 EDT
%%% Non-responding authors or those who said NO are commented out.
%%% ====================================================================
%%% Click the RELOAD button on your web browser to see the updated file.
%%% ====================================================================
%%% Use \input{author} to insert this material into your latex file.
%%%%% Force institutions to appear in alphabetical order when typeset.
\affiliation{Aomori University, Aomori}
\affiliation{Budker Institute of Nuclear Physics, Novosibirsk}
\affiliation{Chiba University, Chiba}
\affiliation{Chonnam National University, Kwangju}
\affiliation{Chuo University, Tokyo}
\affiliation{University of Cincinnati, Cincinnati, Ohio 45221}
\affiliation{University of Frankfurt, Frankfurt}
\affiliation{Gyeongsang National University, Chinju}
\affiliation{University of Hawaii, Honolulu, Hawaii 96822}
\affiliation{High Energy Accelerator Research Organization (KEK), Tsukuba}
\affiliation{Hiroshima Institute of Technology, Hiroshima}
\affiliation{Institute of High Energy Physics, Chinese Academy of Sciences, Beijing}
\affiliation{Institute of High Energy Physics, Vienna}
\affiliation{Institute for Theoretical and Experimental Physics, Moscow}
\affiliation{J. Stefan Institute, Ljubljana}
\affiliation{Kanagawa University, Yokohama}
\affiliation{Korea University, Seoul}
\affiliation{Kyoto University, Kyoto}
\affiliation{Kyungpook National University, Taegu}
\affiliation{Swiss Federal Institute of Technology of Lausanne, EPFL, Lausanne}
\affiliation{University of Ljubljana, Ljubljana}
\affiliation{University of Maribor, Maribor}
\affiliation{University of Melbourne, Victoria}
\affiliation{Nagoya University, Nagoya}
\affiliation{Nara Women's University, Nara}
\affiliation{National Central University, Chung-li}
\affiliation{National Kaohsiung Normal University, Kaohsiung}
\affiliation{National United University, Miao Li}
\affiliation{Department of Physics, National Taiwan University, Taipei}
\affiliation{H. Niewodniczanski Institute of Nuclear Physics, Krakow}
\affiliation{Nihon Dental College, Niigata}
\affiliation{Niigata University, Niigata}
\affiliation{Osaka City University, Osaka}
\affiliation{Osaka University, Osaka}
\affiliation{Panjab University, Chandigarh}
\affiliation{Peking University, Beijing}
\affiliation{Princeton University, Princeton, New Jersey 08545}
\affiliation{RIKEN BNL Research Center, Upton, New York 11973}
\affiliation{Saga University, Saga}
\affiliation{University of Science and Technology of China, Hefei}
\affiliation{Seoul National University, Seoul}
\affiliation{Sungkyunkwan University, Suwon}
\affiliation{University of Sydney, Sydney NSW}
\affiliation{Tata Institute of Fundamental Research, Bombay}
\affiliation{Toho University, Funabashi}
\affiliation{Tohoku Gakuin University, Tagajo}
\affiliation{Tohoku University, Sendai}
\affiliation{Department of Physics, University of Tokyo, Tokyo}
\affiliation{Tokyo Institute of Technology, Tokyo}
\affiliation{Tokyo Metropolitan University, Tokyo}
\affiliation{Tokyo University of Agriculture and Technology, Tokyo}
\affiliation{Toyama National College of Maritime Technology, Toyama}
\affiliation{University of Tsukuba, Tsukuba}
\affiliation{Utkal University, Bhubaneswer}
\affiliation{Virginia Polytechnic Institute and State University, Blacksburg, Virginia 24061}
\affiliation{Yonsei University, Seoul}
  \author{K.~Abe}\affiliation{High Energy Accelerator Research Organization (KEK), Tsukuba} % KEK
  \author{K.~Abe}\affiliation{Tohoku Gakuin University, Tagajo} % TohokuGakuin
  \author{N.~Abe}\affiliation{Tokyo Institute of Technology, Tokyo} % TIT
  \author{I.~Adachi}\affiliation{High Energy Accelerator Research Organization (KEK), Tsukuba} % KEK
  \author{H.~Aihara}\affiliation{Department of Physics, University of Tokyo, Tokyo} % Tokyo
  \author{M.~Akatsu}\affiliation{Nagoya University, Nagoya} % Nagoya
  \author{Y.~Asano}\affiliation{University of Tsukuba, Tsukuba} % Tsukuba
  \author{T.~Aso}\affiliation{Toyama National College of Maritime Technology, Toyama} % Toyama
  \author{V.~Aulchenko}\affiliation{Budker Institute of Nuclear Physics, Novosibirsk} % BINP
  \author{T.~Aushev}\affiliation{Institute for Theoretical and Experimental Physics, Moscow} % ITEP
% ...
% Lausanne
\collaboration{The Belle Collaboration}
\noaffiliation

\begin{abstract}
We report a  measurement of the branching fractions  
in the $B^0 \rightarrow J/\psi \pi^+\pi^-$ decay based
on a $140\,{\rm fb}^{-1}$ data sample 
collected at the $\Upsilon(4S)$ energy
with the Belle detector at the KEKB $e^+ e^-$ collider. 
Charged pion pairs are found to arise mainly from  
$\rho^0$ and $f_2$ mesons; an upper limit for the
non-resonant contribution is also set.
The following branching fractions are obtained: 
${\cal B}(B^0 \rightarrow J/\psi \rho^0)$
%=$(2.8 \pm 0.3\mbox{stat.}) \pm 0.3(\mbox{syst.}))\times 10^{-5}$,
=$(2.8 \pm 0.3 \pm 0.3)\times 10^{-5}$,
${\cal B}(B^0 \rightarrow J/\psi f_2)$
$< 1.5\times 10^{-5} \mbox{(at 90\% C.L.)}$ 
%KM ${\cal B}(B^0 \rightarrow J/\psi f_2)$
%KM =$(9.7 \pm 3.9(\mbox{stat.}) \pm 1.9(\mbox{syst.}))\times 10^{-6}$,
and
${\cal B}(B^0 \rightarrow J/\psi (\pi^+\pi^-)_{\rm non-res.})$
$< 1 \times 10^{-5}$ (at 90\% C.L.).
\end{abstract}

%\pacs{13.65.+i, 13.25.Gv, 14.40.Gx}

\maketitle

%%%% >>>> keep the final version single-spaced
\tighten

{\renewcommand{\thefootnote}{\fnsymbol{footnote}}}
\setcounter{footnote}{0}
%%%%%%%%%%%%%%
% Introduction
%%%%%%%%%%%%%%
The decay $B^0 \to J/\psi \rho^0$~\cite{CC}, which is governed by 
the $b \rightarrow c\bar{c}d$ transition, can exhibit a 
$CP$-violating asymmetry.
Since the tree diagram of this transition has the same weak phase as 
$b \rightarrow c\bar{c}s$, a measurement of indirect $CP$ violation
in this decay provides an alternative estimate of $\sin 2 \phi_1$.
In contrast to the $b \rightarrow c \bar{c} s$ case, however, 
both tree and penguin amplitudes contribute to the 
$b \rightarrow c \bar{c} d$ transition
in the same order of $\sin{\theta_{\rm c}}$.
% the sine of the Cabbibo angle.
Therefore, if penguin or other contributions are substantial, a precision
measurement of the time-dependent $CP$ asymmetry in 
$b\rightarrow c \bar{c}d$ may reveal the values that differ from 
those for $b \rightarrow c \bar{c} s$.
Thus, $B$ decays induced by the $b\rightarrow c \bar{c}d$ transition,
such as $B^0 \rightarrow J/\psi \rho^0$, play an important role in 
probing non-tree diagram contributions.

Since the $\rho^0$ meson has a large width, 
it is necessary to study all the decays of the neutral $B$ meson
that result in a $J/\psi \pi^+ \pi^-$ final state. 
In general, there are two possible contributions to 
the $B^0 \to J/\psi \pi^+ \pi^-$ mode.
One is resonant, when a $\pi^+ \pi^-$ pair arises from some 
resonant state. The $B^0 \to J/\psi \rho^0$ decay mode is 
the largest in this class, although it is  important to search for other 
resonant contributions. The other contribution is non-resonant, 
where a neutral $B$ meson decays directly into a $J/\psi$ and 
$\pi^+\pi^-$ pair. For this process, the $CP$ eigenvalue is unknown; 
therefore, the evaluation of this contribution is important to 
control uncertainties when measuring $CP$ violation in 
$B^0 \to J/\psi \rho^0$ decays. 

%%%%%%%%%%%%%%%%%%%%%%%%%%%%%%%
% Previous work and Data sample
%%%%%%%%%%%%%%%%%%%%%%%%%%%%%%%
The first attempt to measure the branching fraction of 
$B^0 \rightarrow J/\psi \rho^0$ was made by 
the CLEO collaboration who set an upper limit
$\mathcal{B}(B^0 \to J/\psi \rho^0) < 2.5 \times 10^{-4}$ at 90\% C.L.
using 2.39~fb$^{-1}$ of data~\cite{CLEOpsirho}. 
Recently the BaBar collaboration reported their measurement of 
the branching fractions 
$\mathcal{B}(B^0 \rightarrow J/\psi \pi^+\pi^-)=
(4.6 \pm 0.7 \pm 0.6) \times 10^{-5}$ 
and $\mathcal{B}(B^0 \to J/\psi \rho^0) = 
(1.6 \pm 0.6 \pm 0.4) \times 10^{-5}$ 
based on a data sample of 52~fb$^{-1}$~\cite{BaBarpsipipi}.
In this paper, we report a measurement of the branching fractions of 
the neutral $B$ meson decays resulting in the $J/\psi \pi^+\pi^-$ final state
via $B^0 \rightarrow J/\psi \rho^0$ and
$B^0 \to J/\psi f_2$  as well as an upper limit on 
the non-resonant $B^0 \to J/\psi \pi^+\pi^-$ decay mode.
These measurements are based on a $140~{\rm fb}^{-1}$ data sample,
which contains 152 million $B\overline{B}$ pairs, 
collected  with the Belle detector~\cite{Belle} at the KEKB 
asymmetric-energy $e^+e^-$ (3.5 on 8~GeV) collider~\cite{KEKB}
operating at the $\Upsilon(4S)$ resonance.

%------
% Description of Belle detector
%------
The Belle detector is a large-solid-angle magnetic spectrometer.
Closest to the interaction point is 
a three-layer silicon vertex detector (SVD), 
followed by a 50-layer central drift chamber (CDC), 
an array of aerogel threshold \v{C}erenkov counters (ACC), 
a barrel-like arrangement of time-of-flight
scintillation counters (TOF), 
and an electromagnetic calorimeter comprised of CsI(T$l$) crystals (ECL).
These subdetectors are located inside a superconducting solenoid coil 
that provides a 1.5~T magnetic field.  
An iron flux-return located outside of the coil is instrumented to detect 
$K_L^0$ mesons and to identify muons (KLM).  
The detector is described in detail elsewhere~\cite{Belle}.

%%%%%%%%%%%%%%%%%
% Event selection
%%%%%%%%%%%%%%%%%
Hadronic events are selected if they satisfy the following criteria:
at least three reconstructed charged tracks;
a total reconstructed ECL energy in the center of mass (cms) frame
in the range between 0.1$\sqrt{s}$ and 0.8$\sqrt{s}$, where
$\sqrt{s}$ is the total cms energy;
an average ECL cluster energy below 1~GeV;
at least one ECL shower in the region $-0.7 < \cos\theta < 0.9$ in
the laboratory frame;
a total visible energy, which is the sum of charged track momenta
and total ECL energy, exceeding 0.2$\sqrt{s}$,
and a reconstructed primary vertex that is consistent with the known
interaction point.
After imposing these requirements, 
the efficiency for selecting $B$-meson pairs that include
a $J/\psi$ meson is estimated by Monte Carlo (MC) simulation to be 99\%.
To suppress continuum events, we require the event shape variable $R_2$
to be less than 0.5,
where $R_2$ is the ratio of the second to the zeroth
Fox-Wolfram moment~\cite{FWM}.

%---------
% Mbc and DeltaE description
%--------
The $B$ candidate selection is carried out using two observables 
in the rest frame of the $\Upsilon(4S)$ (cms):
the beam-energy constrained mass 
$M_{\rm bc} \equiv \sqrt{ E_{beam}^2 - (\sum \vec{p_i})^2} $ 
and the energy difference $\Delta E \equiv \sum E_i - E_{beam}$,
where $E_{beam}=\sqrt{s}/2$ is the cms beam energy,
and $\vec{p_i}$ and $E_i$ are the cms three-momenta and energies of 
the $B$ meson decay product candidates.  

%-----------
% Reconstruction of $J/\psi$ mesons
%----------
Candidate $J/\psi$ mesons are reconstructed via their decay into oppositely 
charged lepton pairs ($e^+e^-$ or $\mu^+\mu^-$).   
Leptons are selected from 
charged tracks satisfying a cut $|dz|<5\mbox{~cm}$, where $dz$ 
is the track's closest approach to the interaction point along 
the beam direction.
For electron identification, the ratio between the charged track's
momentum and the associated shower energy ($E/p$) is the most powerful
discriminant.
Other information, including the CDC specific ionization, 
the distance between the ECL shower
and the extrapolated track, and the shower shape is also used.
Muons are identified by requiring an association between KLM hits and
an extrapolated track.
Both lepton tracks must be positively identified as such.
In the $e^+e^-$ mode, ECL clusters
that are within 50~mrad of the track's initial momentum vector are
included in the calculation of the invariant mass ($M_{ee(\gamma)}$),
in order to include photons radiated
by electrons/positrons.
The invariant masses of $e^+e^-(\gamma)$ and $\mu^+\mu^-$ combinations 
are required to fall in the ranges  
$-0.15\mbox{~GeV}/c^2 < (M_{ee(\gamma)} - M_{J/\psi} ) 
< +0.036\mbox{~GeV}/c^2$ and
$-0.06\mbox{~GeV}/c^2 < (M_{\mu\mu} - M_{J/\psi}) 
< +0.036\mbox{~GeV}/c^2$, respectively. 
Here $M_{J/\psi}$ denotes the world average of the 
$J/\psi$ mass~\cite{PDG}.
We perform a vertex fit to the lepton pair candidate.
Then a mass constrained fit is applied
to improve the $\Delta E$ and $M_{\rm bc}$ resolutions of 
the selected $B$ meson candidates. 

%------
% pion candidate selection
%------
Information from the ACC, TOF and CDC is combined 
into a likelihood ratio for kaon/pion separation; we then impose a cut 
to reject kaons.
The efficiency for pions is more than 85\% while the kaon fake rate 
is 10\%. Charged pion candidates are also rejected if they are 
positively identified as leptons.
Using pion candidates with $|dz|<$ 5~cm, 
$\pi^+\pi^-$ pairs are formed.
We apply a vertex fit to $\pi^+\pi^-$ pairs and require 
the distance between the reconstructed vertices 
of the $J/\psi$ and the $\pi^+\pi^-$ pair to be less than 3 mm
in order to suppress $B^0 \to J/\psi K^0_S, K^0_S \to \pi^+ \pi^-$ 
events and 
backgrounds due to accidentally formed pion pairs.

\begin{figure}[htb]
\includegraphics[width=0.7\textwidth]{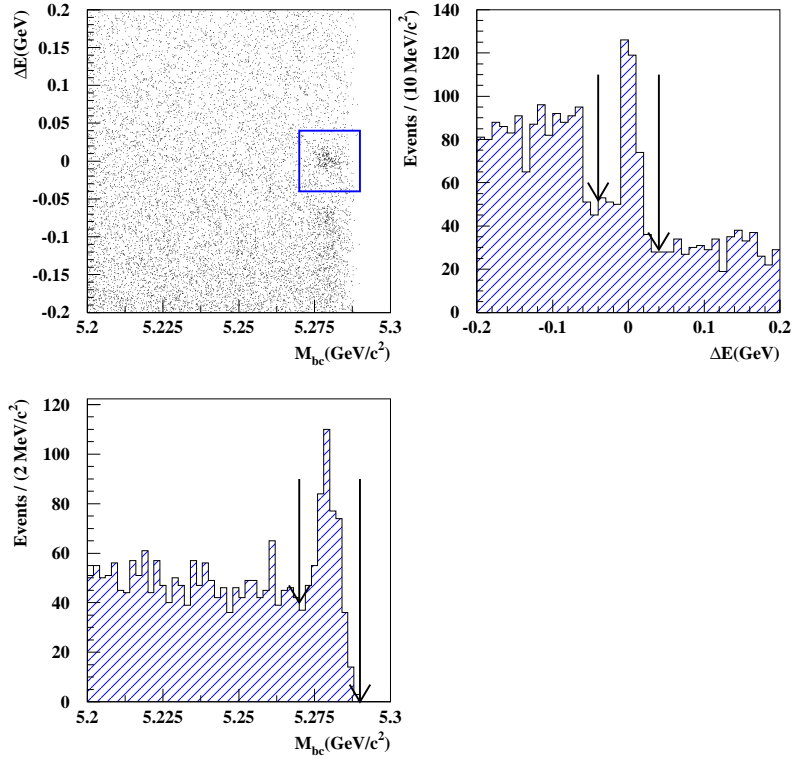}
\caption{(a) Distribution of $\Delta E$ versus $M_{\rm bc}$ for
$B^0 \rightarrow J/\psi \pi^+\pi^-$ candidates in data.
(b) Projection in $\Delta E$ for 
$5.270\mbox{~GeV}/c^2 < M_{\rm bc} < 5.290\mbox{~GeV}/c^2$
and (c) Projection in $M_{\rm bc}$ for 
$-0.04\mbox{~GeV} < \Delta E < 0.04\mbox{~GeV}$.
The arrows in (b) and (c) indicate the boundaries of the signal box.}
\label{mbcdeltae}
\end{figure}

$J/\psi$ candidates and $\pi^+\pi^-$ pairs are combined to select 
$B$ candidates.
The distribution of the candidates in $\Delta E$ and $M_{\rm bc}$ 
is shown in Fig. \ref{mbcdeltae}.
The signal box to select  candidate events is defined as
$5.270\mbox{~GeV}/c^2 < M_{\rm bc} < 5.290\mbox{~GeV}/c^2$ 
and $-0.04\mbox{~GeV} < \Delta E < 0.04\mbox{~GeV}$. 
%KM According to MC, the detection efficiency is estimated to be 27.8\%,
%KM Using a MC data set of 
%KM $B^0 \rightarrow J/\psi \pi^+ \pi^-$ phase space decay,
Using a MC data set of $B^0 \rightarrow J/\psi \pi^+ \pi^-$ decays 
distributed uniformly in phase space,
the detection efficiency is estimated to be 27.8\%,
where a difference of the pion identification in efficiency between
MC and data is taken into account: the MC efficiency is 2.8\% higher per
pion track than in data.
After applying all selection criteria, there are 537 candidates 
remaining in the signal box.

%-----------------
% Mpipi fit description
%-----------------
A binned maximum likelihood fit is performed to the distribution of 
the two pion invariant mass ($M_{\pi^+\pi^-}$) of the selected events 
to determine various contributions to 
$B^0 \rightarrow J/\psi \pi^+\pi^-$ events.
Five types of events are  considered: 
(i) $B^0 \rightarrow J/\psi \rho^0$; 
(ii) $B^0 \rightarrow J/\psi f_2$; 
(iii) $B^0 \rightarrow J/\psi \pi^+\pi^-$ (non-resonant signal);
(iv) $B^0 \rightarrow J/\psi K^0_S (K^0_S \rightarrow \pi^+\pi^-)$; 
(v) background.
A probability density function (PDF) is constructed for 
each of these five cases.  
The $B^0 \rightarrow J/\psi K^0_S$ mode is not considered as a 
signal while determining the branching fractions
for the decay modes contributing to $B^0 \rightarrow J/\psi \pi^+\pi^-$.

The PDF used to model the $B^0 \rightarrow J/\psi\rho^0$ mode is 
a relativistic $P$-wave Breit-Wigner function~\cite{PisutRoos}:\\
\begin{eqnarray}
F_{\rho}(M_{\pi^+\pi^-})
&=&  \frac{ M_{\pi^+\pi^-} \Gamma(M_{\pi^+\pi^-}) P^{2L_{\rm eff}+1}}
 {(M_{\rho}^2-M_{\pi^+\pi^-}^2)^2 + M_{\rho}^2\Gamma(M_{\pi^+\pi^-})^2}
\end{eqnarray}
where
\begin{eqnarray}
\Gamma(M_{\pi^+\pi^-}) &=& \Gamma_0 (\frac{q}{q_0})^{2 l + 1} 
(\frac{M_{\rho}}{M_{\pi^+\pi^-}}) (\frac{1 + R^2 q_0^2}{1 + R^2 q^2}).
\end{eqnarray}
Here, $q(M_{\pi^+\pi^-})$ is the pion momentum in the di-pion rest frame, 
with $q_0 = q(M_{\rho})$;  
$P$ is the $J/\psi$ momentum in the $B^0$ rest frame;  
$l=1$ is the $\rho^0$ meson's spin;
$M_{\rho} = 775.8~\mbox{MeV}/c^2$ and
$\Gamma_0 = 146.4~\mbox{MeV}/c^2$.
Also, $L_{\rm eff}$ is the effective orbital angular momentum between 
the $J/\psi$ and the $\rho^0$, which can take any value 
between 0 and 2,
and $R$ is the Blatt-Weisskopf barrier-factor radius~\cite{Blatt}.  
The fit is performed with $L_{\rm eff}$ and $R$ equal to 1
and 0.5~fm, respectively.
 
The PDF for the $B^0 \rightarrow J/\psi f_2$ mode is a 
relativistic $D$-wave Breit-Wigner function that is obtained by 
replacing the appropriate parameters in the PDF for 
$B^0 \rightarrow J/\psi \rho^0$: $l=2$, 
$M_{\rm f_2} = 1.285~\mbox{GeV}/c^2$ and 
$\Gamma_0 = 184.3~\mbox{MeV}/c^2$.

The PDF used to model the $B^0 \rightarrow J/\psi K^0_S$ mode is 
a single Gaussian function with the mass and width fixed to the values 
obtained by fitting a $K^0_S \rightarrow \pi^+\pi^-$ invariant
mass distribution in $B^0 \rightarrow J/\psi K^0_S$ MC events.
%KM in hadronic events.

%KM The PDF for the 
%KM $B^0 \rightarrow J/\psi \pi^+\pi^-\mbox{non-resonant signal}$ 
%KM is an empirical second order polynomial function compatible with
%KM both lower and upper kinematic limits of $M_{\pi^+\pi^-}$.
The PDF for the 
$B^0 \rightarrow J/\psi \pi^+\pi^-\mbox{non-resonant signal}$ 
is a parameterized $M_{\pi^+\pi^-}$ distribution of 
the $B^0 \rightarrow J/\psi \pi^+ \pi^-$ phase space decay model.

\begin{figure}[htb]
\includegraphics[width=1.0\textwidth]{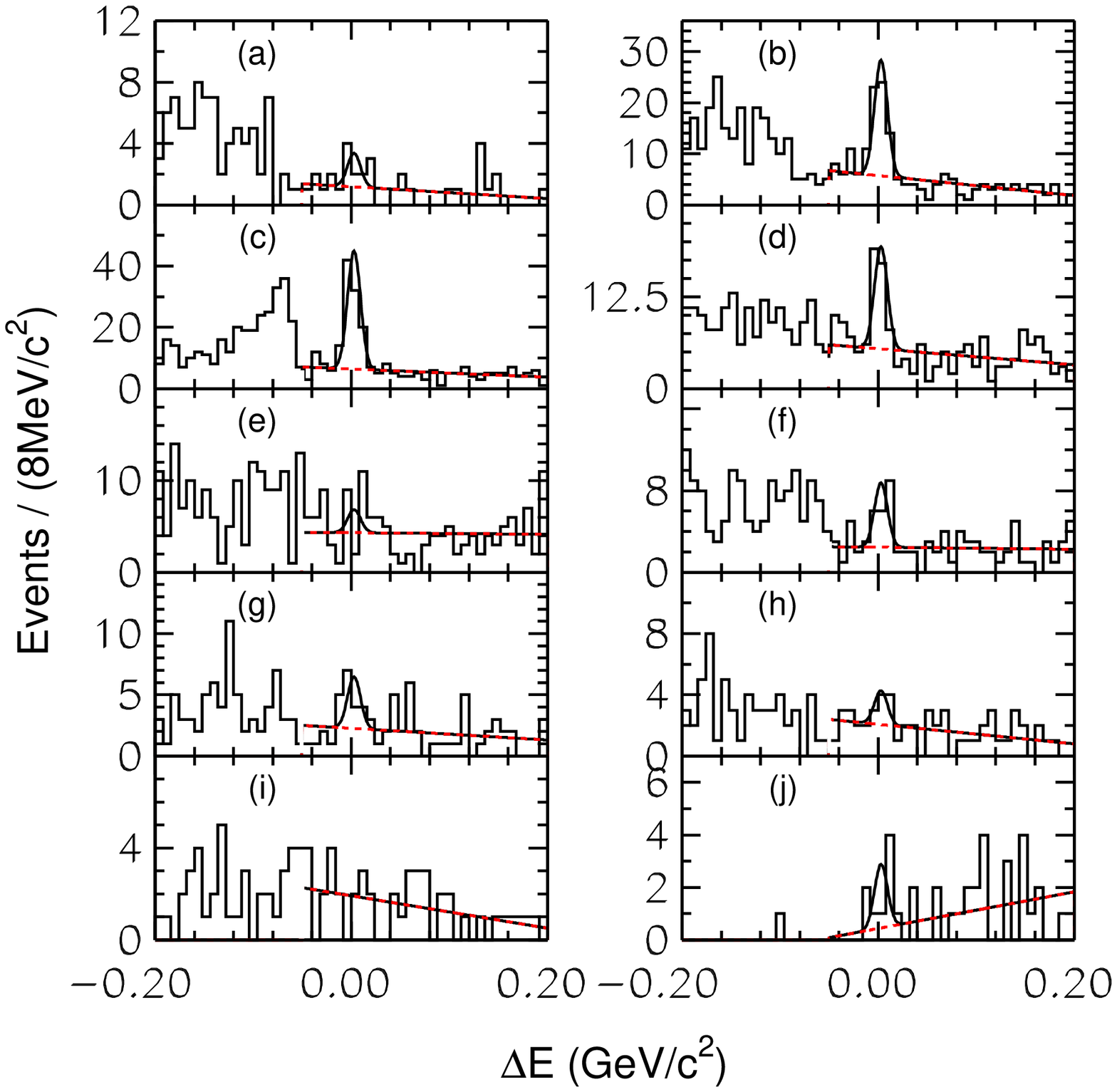}
\caption{The  $\Delta E$  distribution  for
$B^0 \rightarrow J/\psi \pi^+\pi^-$ candidates in data
in different $M_{\pi^+\pi^-}$ regions with the requirement
$5.270~\mbox{GeV}/c^2 < M_{\rm bc} < 5.290~\mbox{GeV}/c^2$.
In each plot the $M_{\pi^+\pi^-}$ range is 0.2 GeV/$c^2$:
from (a) $0.25~\mbox{GeV}/c^2<M_{\pi^+\pi^-}<0.45~\mbox{GeV}/c^2$ to 
%KM (b) $0.45<M_{\pi^+\pi^-}<0.65\mbox{GeV}/c^2$,
%KM (c) $0.65<M_{\pi^+\pi^-}<0.85\mbox{GeV}/c^2$,
%KM (d) $0.85<M_{\pi^+\pi^-}<1.05\mbox{GeV}/c^2$,
%KM (e) $1.05<M_{\pi^+\pi^-}<1.25\mbox{GeV}/c^2$,
%KM (f) $1.25<M_{\pi^+\pi^-}<1.45\mbox{GeV}/c^2$,
%KM (g) $1.45<M_{\pi^+\pi^-}<1.65\mbox{GeV}/c^2$,
%KM (h) $1.65<M_{\pi^+\pi^-}<1.85\mbox{GeV}/c^2$,
%KM (i) $0.85<M_{\pi^+\pi^-}<2.05\mbox{GeV}/c^2$ and
(j) $1.05~\mbox{GeV}/c^2<M_{\pi^+\pi^-}<1.25~\mbox{GeV}/c^2$ 
in the alphabetic order.
The combinatorial background is estimated by fitting this distribution
with a Gaussian for a signal (the solid curve) and 
a first order polynomial for background (the dashed line).
}
\label{deltaEfit}
\end{figure}

The $\Delta E$ distribution is used to determine the PDF for 
the background $M_{\pi^+\pi^-}$ shape.
According to the MC study, background modes do not give a peak
in the $\Delta E$ distribution.
Therefore, we estimate background by fitting the $\Delta E$ distribution 
with a first order polynomial (for background) 
and a Gaussian (for signals and $K^0_S$; $B^0 \to J/\psi \pi^+ \pi^-$ 
final state). 
In order to estimate the background as a function of $M_{\pi^+\pi^-}$, 
we subdivide the candidate event sample 
%according to $M_{\pi^+\pi^-}$ distribution in every 0.2 GeV/$c^2$ bin 
in 0.2 GeV/$c^2$ $M_{\pi^+\pi^-}$   bins 
from 0.25 GeV/$c^2$ to 2.25 GeV/$c^2$.
Then a fit to the $\Delta E$ distribution in each 
$M_{\pi^+\pi^-}$ bin is performed.
%KM modified as Simon suggested. 20040817.
%KM Here, we consider 
%KM the $B^0 \rightarrow J/\psi K^{*0}, K^{*0} \rightarrow K^+ \pi^-$ decay
%KM since it is one of the major background sources when the charged kaon 
%KM is misidentified as pion. 
One of the major background sources here is the 
$B^0 \rightarrow J/\psi K^{*0}, K^{*0} \rightarrow K^+ \pi^-$ decay,
when a charged kaon is misidentified as a pion.
Because of the difference between the kaon and pion masses, 
this background mainly populates a region around 
$\Delta E \approx -0.1~\mbox{GeV}$.
%KM Since it is not easy to parameterize  
%KM $B^0 \rightarrow J/\psi K^{*0}, K^{*0} \rightarrow K^+ \pi^-$ $\Delta E$ 
%KM distribution, 
By choosing the region of 
$-0.05~\mbox{GeV} < \Delta E < 0.2~\mbox{GeV}$ for our
fits  we avoid this contribution.
The signal width and mean value are fixed according to signal MC 
expectations.
Fig.~\ref{deltaEfit} shows the results of the fits.

We fit the obtained background with a fourth order polynomial as a 
function of $M_{\pi^+\pi^-}$.
Since this PDF is determined from data in a  model independent
way, we fix it in the fit.

\begin{figure}[htb]
\includegraphics[width=0.7\textwidth]{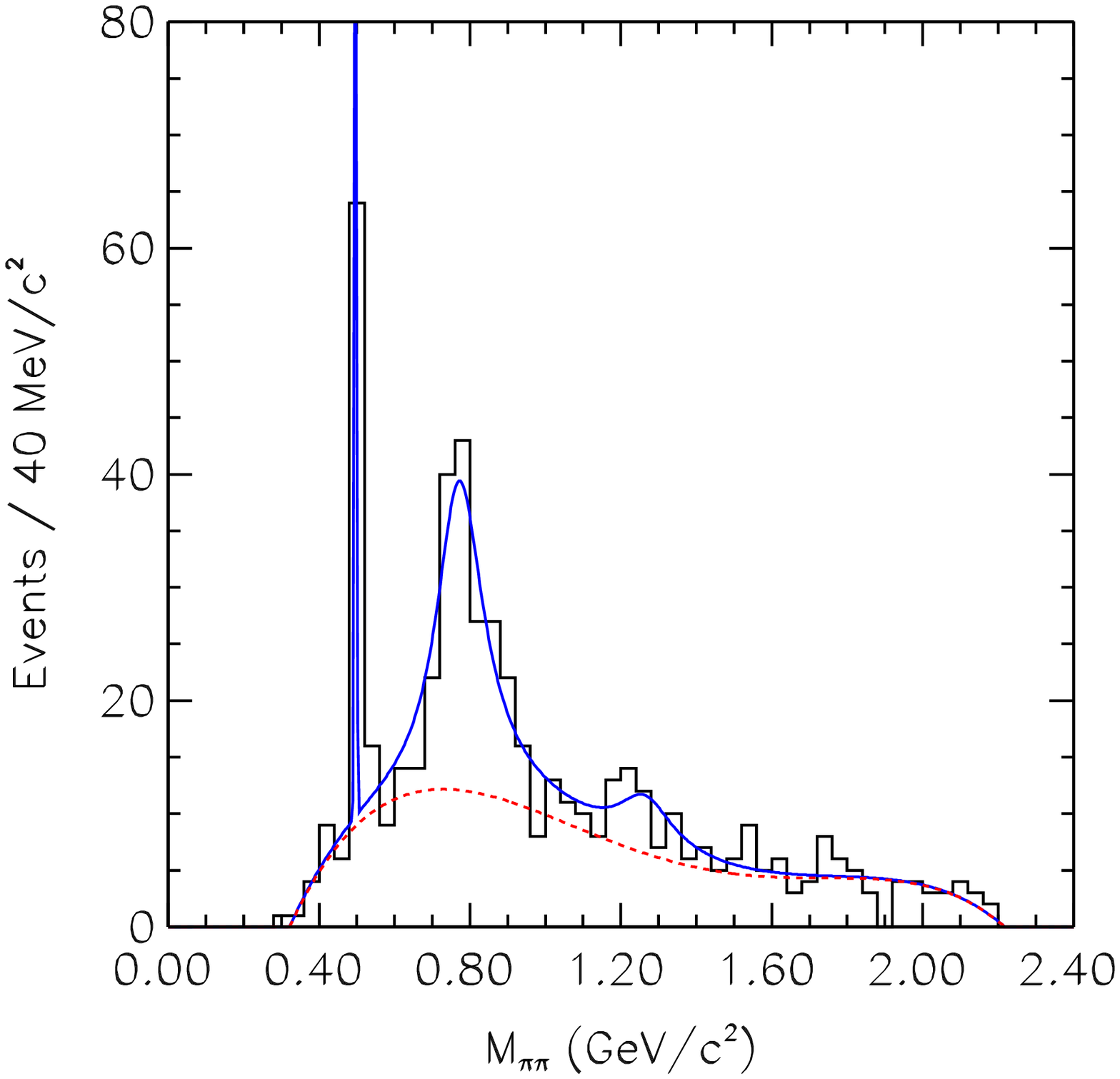}
\caption{The distribution of $M_{\pi^+\pi^-}$ for
$B^0 \rightarrow J/\psi \pi^+\pi^-$ candidates in data.
The solid line shows the fitted results by taking the $K^0_S$, $\rho^0$, 
$f_2$ and background contributions into account.
The dashed line is background only.
}
\label{mpipifit}
\end{figure}

%--------------
% Results of the fit (signal yield)
%-------------
In order to find the probability of the 
$B^0 \rightarrow J/\psi f_2$ and 
non-resonant $B^0 \rightarrow J/\psi \pi^+\pi^-$ decays, 
the fit is carried out in three different cases: 
(i) PDFs for $B^0 \to J/\psi \rho^0$, $B^0 \to J/\psi K_S$ and 
background are summed, 
(ii) the $B^0 \to J/\psi f_2$ contribution is additionally 
taken into account, and
(iii) adding the non-resonant $B^0 \rightarrow J/\psi \pi^+ \pi^-$ 
contribution to case (ii).
The optimal likelihood values, ${\cal L}_{\rm i}$, ${\cal L}_{\rm ii}$ 
and ${\cal L}_{\rm iii}$ for the cases (i), (ii) and (iii), 
respectively, are used to obtain a statistical significance: 
$-2 \ln {\cal L}_{\rm i} = 42.4$, $-2 \ln {\cal L}_{\rm ii} = 33.7$ 
and $-2 \ln {\cal L}_{\rm iii} = 33.7$.
By comparing cases (i) and (ii), a statistical significance of 
$B^0 \rightarrow J/\psi f_2$ is found to be $2.9 \sigma$.
As shown in Fig. \ref{mpipifit}, which is the fit result in case (ii), 
the $M_{\pi^+\pi^-}$ spectrum is almost saturated by the $K_S$, $\rho^0$ and 
$f_2$ contributions, and the non-resonant 
$B^0 \rightarrow J/\psi \pi^+ \pi^-$ signal is found 
to be small.
As a result of the fit in case (ii), the signal yields are obtained to 
be 139.8$\pm$17.0 for $B^0 \rightarrow J/\psi \rho^0$ and 
27.6$\pm$11.0 for $B^0 \rightarrow J/\psi f_2$.
To get a conservative upper limit for the non-resonant 
$B^0 \to J/\psi \pi^+ \pi^-$ contribution, the background PDF 
normalization is reduced by $1 \sigma$ and another fit to 
the $M_{\pi^+\pi^-}$ distribution is performed.
The resulting yield is 15.1 $\pm$ 20.1 events.
Taking systematic uncertainties listed in Table \ref{sys1}, 
and using the Feldman-Cousins approach \cite{FC}, an upper limit
is obtained to be 
${\cal B}(B^0 \rightarrow J/\psi (\pi^+ \pi^-)_{\rm non-res.})
 < 1 \times 10^{-5}$ at 90\% C.L.

%------------
% systematic errors
%------------
For the $B^0 \rightarrow J/\psi \rho^0$ and 
$B^0 \rightarrow J/\psi f_2$ modes,
the following systematic uncertainties are taken into account:
uncertainty of the track detection efficiency;
background estimations;
Breit-Wigner modeling,
uncertainty of the relevant branching fractions;
statistical uncertainties of the signal MC events; and 
uncertainty on the number of $B \overline{B}$ events.

The tracking efficiency is estimated by MC and the control samples of
partially reconstructed $D^*$ mesons, and 
its uncertainty is estimated to be 1.2\% per track.
The lepton identification efficiency is obtained from the ratio between 
the yields of single and double-tagged $J/\psi$ mesons, 
and its uncertainty is found to be 1.9\% per track.
An uncertainty of high-momentum charged pion identification is 
estimated from $D^{*+} \rightarrow D^0(\rightarrow K^- \pi^+) \pi^+$ 
decays to be 1.7\% per pion pair.
The normalization of the background PDF is changed by $\pm 1 \sigma$ 
and a fit to $M_{\pi\pi}$ is performed again.
Then the signal yield change is taken into account as a systematic error
due to background estimation.
An uncertainty due to $L_{\rm eff}$ and $R$ in the Breit-Wigner PDFs
is estimated by the fit that is carried out changing $L_{\rm eff}$ 
over its allower range.
Two different values of the 
Blatt-Weisskopf barrier-factor radius 
$R$ (0.5 and 1.0 fm) are also compared. 
The maximum variation of the signal yield of
each mode is assigned as a systematic error.
Using a large MC sample of $B^0 \rightarrow J/\psi \pi^+ \pi^-$ phase space
decays, the detection efficiency is obtained as a 
function of $M_{\pi^+\pi^-}$ ,
%KM ($\varepsilon(M_{\pi^+\pi^-})$), 
and its product by the Breit-Wigner PDFs for $\rho^0$ and $f_2$ mesons
is integrated. The resulting values are compared to the detection 
efficiency obtained as an average over the whole kinematical region, and
the difference is assigned as a systematic error due to 
the signal MC modeling.

All other investigated systematic uncertainties are small compared to those
described above. We list
all the estimated systematic uncertainties for each mode 
in Table~\ref{sys1}. 

\begin{table}[htb]
\caption{Systematic uncertainties for each decay mode. 
Note that the uncertainty of background estimation for 
non-resonant $J/\psi \pi^+\pi^-$ is already taken into account 
to obtain a yield.}
\label{sys1}
%KM \begin{tabular}
%KM {@{\hspace{0.5cm}}l@{\hspace{0.5cm}}||@{\hspace{0.5cm}}c@{\hspace{0.5cm}}}
\begin{tabular}
{@{\hspace{0.5cm}}l@{\hspace{0.5cm}}||@{\hspace{0.5cm}}c@{\hspace{0.5cm}}|
@{\hspace{0.5cm}}c@{\hspace{0.5cm}}|@{\hspace{0.5cm}}c@{\hspace{0.5cm}}}
\hline \hline
Mode & $J/\psi \rho^0$ & $J/\psi f_2$ & $J/\psi (\pi^+\pi^-)_{\rm non-res.}$ \\
\hline
Track reconstruction & $\pm 4.8$ \% & $\pm 4.8$ \% & $\pm 4.8$ \%  \\
Lepton-ID & $\pm 3.8$ \% & $\pm 3.8$ \% & $\pm 3.8$ \% \\
$\pi^{\pm}$-ID & $\pm 1.7$ \% & $\pm 1.7$ \% & $\pm 1.7$ \% \\
Background estimation &  $\pm 5.1$ \% & $\pm 15.4$ \% & - \\
Breit-Wigner modeling &  $\pm 2.7$ \% & $\pm 10.9$ \% & - \\
Signal MC modeling &  $\pm 2.7$ \% & $\pm 2.4$ \% & -  \\
MC statistics &  $\pm 0.7$ \% & $\pm 0.7$ \% & $\pm 0.7$ \% \\
${\cal B}(J/\psi \rightarrow \ell^+ \ell^-)$ 
              &  $\pm 1.7$ \% & $\pm 1.7$ \% & $\pm 1.7$ \% \\
${\cal B}(f_2 \rightarrow \pi^+ \pi^-)$ 
              &  - & $\pm 2.9$ \% & - \\
The number of $B\overline{B}$ 
              &  $\pm 0.4$ \% & $\pm 0.4$ \% & $\pm 0.4$ \% \\
\hline
Total & $\pm 9.2$ \% & $\pm 20.3$ \% & $\pm 6.6$ \%\\
\hline \hline
\end{tabular}
\end{table}

%----------
% conclusions
%----------
Based on a large sample of $B\overline{B}$ pairs,
we obtained the following preliminary branching fractions 
for $B$ decays resulting in the $J/\psi \pi^+\pi^-$ final state:
${\cal B}(B^0 \rightarrow J/\psi\rho^0) 
= (2.8\pm0.3(\mbox{stat.})\pm0.3(\mbox{syst.}))\times 10^{-5}$ and 
${\cal B}(B^0 \rightarrow J/\psi f_2)
=(9.8\pm3.9(\mbox{stat.})\pm2.0(\mbox{syst.}))\times 10^{-6}$.
Since a statistical significance for the latter is $2.9 \sigma$ only, 
we also set an upper limit 
${\cal B}(B^0 \rightarrow J/\psi f_2)<1.5\times10^{-5}$ at the 90\% C.L.;
we also find
${\cal B}(B^0 \rightarrow J/\psi (\pi^+ \pi^-)_{\rm non-res.})
<1 \times 10^{-5}$ at the 90\% C.L.
%KM added 20040817 as Simon suggested.
The obtained value of ${\cal B}(B^0 \rightarrow J/\psi \rho^0)$
is consistent with that of BaBar~\cite{BaBarpsipipi} and is more precise. \\
%\section*{Acknowledgments}
%***** Acknowledgments *****
% Please paste this acknowledgement into your latex file. 
 %----------- Long version, for most papers ----------- 
We thank the KEKB group for the excellent operation of the
accelerator, the KEK Cryogenics group for the efficient
operation of the solenoid, and the KEK computer group and
the National Institute of Informatics for valuable computing
and Super-SINET network support. We acknowledge support from
the Ministry of Education, Culture, Sports, Science, and
Technology of Japan and the Japan Society for the Promotion
of Science; the Australian Research Council and the
Australian Department of Education, Science and Training;
the National Science Foundation of China under contract
No.~10175071; the Department of Science and Technology of
India; the BK21 program of the Ministry of Education of
Korea and the CHEP SRC program of the Korea Science and
Engineering Foundation; the Polish State Committee for
Scientific Research under contract No.~2P03B 01324; the
Ministry of Science and Technology of the Russian
Federation; the Ministry of Education, Science and Sport of
the Republic of Slovenia; the National Science Council and
the Ministry of Education of Taiwan; and the U.S.\
Department of Energy.

%-------- Short version, if necessary, for PRL -----------
% currently commented out
%We thank the KEKB group for the excellent operation of the
%accelerator, the KEK Cryogenics group for the efficient
%operation of the solenoid, and the KEK computer group and
%the NII for valuable computing and Super-SINET network
%support.  We acknowledge support from MEXT and JSPS (Japan);
%ARC and DEST (Australia); NSFC (contract No.~10175071,
%China); DST (India); the BK21 program of MOEHRD and the CHEP
%SRC program of KOSEF (Korea); KBN (contract No.~2P03B 01324,
%Poland); MIST (Russia); MESS (Slovenia); NSC and MOE
%(Taiwan); and DOE (USA).

%%%%%%%%%%%%%%%%%%%%%%%%%%%
% Reference
%%%%%%%%%%%%%%%%%%%%%%%%%%%

%


\begin{thebibliography}{99}
\bibitem{CC}
Throughout this paper, 
the inclusion of the charge conjugate decay modes is implied
unless otherwise stated.
\end{thebibliography}
\end{document}